\documentclass[showpacs,prl,superscriptaddress,twocolumn]{revtex4-1}
\usepackage[table]{xcolor}
\usepackage{textcomp}
\usepackage{amsmath}
\usepackage{amssymb}
\usepackage{graphicx} 
\usepackage{pgf}
\usepackage{hyperref}
\usepackage{epstopdf}
\usepackage{braket}
\usepackage{mathtools}

\def\fref#1{Fig.~\ref{#1}}

\begin{document}

\title{Dynamical Phase Diagram of a Quantum Ising Chain with Long Range Interactions}

\author{Giulia Piccitto}
\affiliation{SISSA, Via Bonomea 265, I-34135 Trieste, Italy}

\author{Bojan \v Zunkovi\v c}
\affiliation{Department of Physics, Faculty of Mathematics and Physics, University of Ljubljana, Jadranska 19, 1000 Ljubljana, Slovenia}

\author{Alessandro Silva}
\affiliation{SISSA, Via Bonomea 265, I-34135 Trieste, Italy}

\begin{abstract}
    We investigate the effect of short-range correlations on the dynamical phase diagram of quantum many-body systems with long-range interactions. Focusing on Ising spin chains with power-law decaying interactions and accounting for short-range correlations by a \textit{cluster mean field} theory we show that short-range correlations are responsible for the emergence of a chaotic dynamical region. Analyzing the fine details of the phase diagram, we show that the resulting chaotic dynamics bears close analogies with that of a tossed coin.
\end{abstract}
\pacs{05.30.Rt, 64.60.Ht, 75.10.Jm}

\maketitle

Interactions have a crucial role in qualitatively determining the dynamics of a quantum many-body system~\cite{noneq:1}. Setting aside the well known relation between integrability and lack of thermalization~\cite{quench:2,quench:3,quench:4,quench:5} the range of interactions determines
the propagation of energy and information in the system~\cite{spread:2,spread:4}, as well as the presence or absence of long-range order at finite temperature or in dynamical situations~\cite{biroli:dqpt,gambassi:dqpt,dqpt:1}. In particular, even after an abrupt change of the transverse field (a \it quantum quench\rm), spin chains with long-range interactions may sustain an order parameter whose vanishing as a function of the quench parameter signals dynamical phase transitions ~\cite{dqpt:1,sperimentale:1}. The latter have been extensively studied both theoretically
and experimentally in trapped ions~\cite{dqpt:1,sperimentale:1,dqpt:2,dqpt:3,dqpt:4}. 

Unlike static transitions, we associate dynamical phase transitions to a qualitative change in the symmetry of the order parameter dynamics~\cite{biroli:dqpt,gambassi:dqpt,dqpt:1}. 
In particular, in Ising systems with infinite range interactions the critical point is characterized by a fragile critical orbit~\cite{lmg:1,dqpt:1}, therefore, the critical point can be easily disrupted by fluctuations due to residual short-range interactions, fanning out in a chaotic phase where predicting the asymptotic sign of the magnetization becomes increasingly difficult~\cite{dqpt:4}. The sensitivity of the asymptotic magnetization to the system parameters is qualitatively reminiscent of the behavior of the ultimate aleatory process, the coin toss~\cite{coin:1}.  The dynamics of a tossed coin depends crucially on how the energy provided to
it is dissipated ~\cite{coin:1}: If the coin falls on sand dissipating all the energy in one shot the result (head or tails) will depend smoothly on the initial conditions. 
If instead, it falls on concrete, dissipating its energy in more and more jumps, the outcome will tend towards true randomness. 

\begin{figure}[h!]
    \centering
    \includegraphics[width=0.45\textwidth]{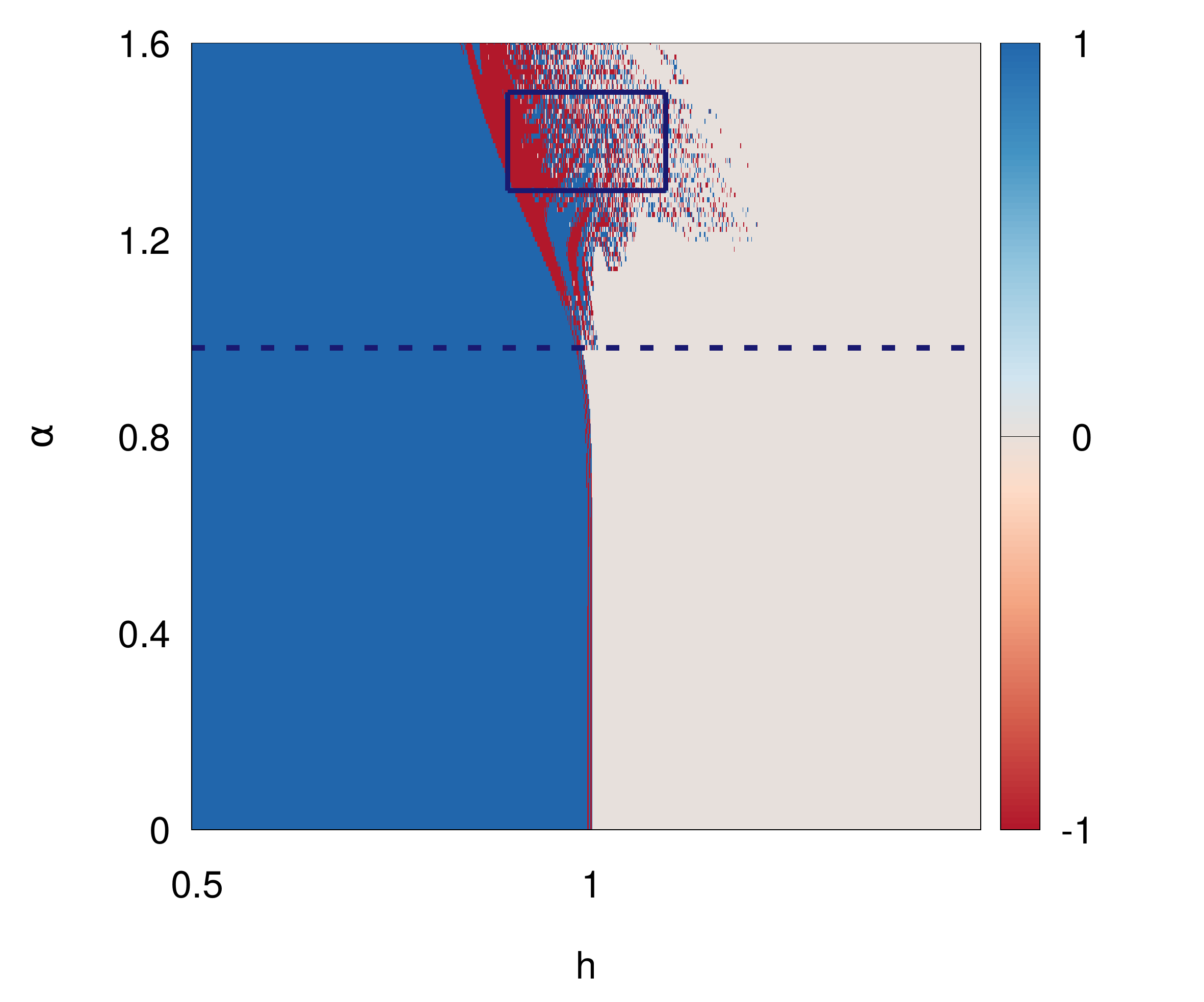}
	\caption{(Color online). The dynamical phase diagram for the long range Ising model (J=1) obtained by CMFT with $\ell=5$ for a quench of the transverse field from $h_{in}=0$ to $h$: the time-averaged longitudinal magnetization (up to a time $T = 100$)  vs. $\alpha$ and $h$. Notice that, as soon as  $\alpha > 1$ (dashed line), the dynamical  phase transition fans out in a critical region,  where the asymptotic magnetization becomes very sensitive to the system parameters. The blue square denotes the region detailed in \fref{Chaos1}.
}
    \label{ddf} 
\end{figure}

The purpose of this paper is to explore in full detail the analogy between a coin toss and the dynamics of an Ising spin chain close to a dynamical critical point. We will focus on the experimentally relevant case of chains with interactions decaying as $1/r^{\alpha}$.
In order to investigate the role of short-range correlations we will describe the dynamics of the system with a \it cluster mean field \rm  theory (CMFT)~\cite{fazio:cluster} : we divide the system into clusters of size $\ell$ whose dynamics is solved exactly in the mean field 
generated by others. The versatility of this method allows us to study the phase diagram for different $\alpha$, showing that the chaotic phase appears only for $\alpha>1$ (see ~\fref{ddf}). Besides, by deriving the phase diagram for different $\ell$, starting from $\ell=1$ (mean field theory), we will show that the key ingredient to obtain a chaotic phase are \it short-range correlations \rm, which are known to influence qualitatively phase diagrams also in other contexts~\cite{fazio:cluster}. Finally, an analysis of the fine details of the phase diagram will
show that the analogy with a coin toss can be taken further: the asymptotic state attained by the spin-chain is more and more sensitive to the initial conditions the higher is the energy provided to the chain (and the longer it takes to dissipate it).

\emph{The model---} 
We study the 1D-transverse field Ising model with power-law decaying interactions, described by the following Hamiltonian
\begin{equation}
    H = -\sum_{i> j}^L J_{ij} \sigma^z_i \sigma^z_j - h \sum_i^L \sigma^x_i,
    \label{model}
\end{equation}
where $L$ is the number of sites, $\sigma^\mu_i$ are the Pauli matrices acting on site $i$, $J_{ij} = \frac{1}{\mathcal{N}_{\alpha,L}} \frac{J}{|i-j|^\alpha}$,
and $\mathcal{N}_{\alpha,L}
 = \sum_{n=1}^L |n|^{-\alpha}$ is the usual Kac normalization.

The model with $\alpha = 0$, named after  Lipkin, Meshkov, and Glick (LMG), describes the dynamics of a large collective spin. The associated phase space in the mean-field limit,
which is exact for $L \to \infty$, reduces to the surface of a Bloch sphere \cite{lmg:1, lmg:2}. 
At equilibrium, this model exhibits a transverse field driven quantum phase transition at $h_c = 2J$ from a ferromagnetic phase to a paramagnetic phase. 
The nonequilibrium dynamics after a quench can be characterized by using as order parameter the time-averaged longitudinal magnetization $m^z = \lim_{T \to \infty} \frac{1}{T} \int_0^T \braket{ \sigma^z(t)} dt$. Starting from a fully polarized initial state, the system shows a dynamical quantum phase transition at $h_c = J$. 
In the thermodynamic limit, and for $0 < \alpha < 1$, the long-range Ising chain is expected to be equivalent to the LMG model \cite{mori}. 
In turn, as soon as the power law exponent crosses the critical value $\alpha =1$, we expect fluctuations to start playing an essential role in the collective dynamics of the system. 

\emph{Cluster mean field approach---} 
In order to describe systematically the effect of fluctuations for $\alpha > 1$ we divide the system into a set of clusters of $\ell$ spins.
We account for short-range correlations by solving the dynamics of each cluster exactly while treating the interaction among different clusters at the mean-field level. 
Formally, this amounts to rewriting the Hamiltonian as a sum of two terms $H = H_{\text{cl}} + H_{\text{out}}$
\begin{equation}
\begin{aligned}
    &H_{\text{cl}} = -\sum_\mu^{N_{\text{cl}}}\left( \sum_{i < j \in \mu} J_{ij} \sigma_i^z     \sigma_j^z - h \sum_{i \in \mu} \sigma^x_i\right)\\
    &H_{\text{out}} =- \sum_{\substack{\mu,\nu \\ \mu \ne \nu}}^{N_\text{cl}} \sum_{\substack{i \in \mu \\j \in \nu}} J_{ij}\sigma_i^z \sigma_j^z,\\ 
\end{aligned}
\end{equation}
where $N_{cl}=L/\ell$, $\mu$ and $\nu$ are cluster indices.
Using the usual mean-field prescription $\sigma_i^z = \bar{m} + \delta$, with $\bar{m} = \frac{1}{\ell}\sum_{i \in \mu}^l \braket{\sigma^z_i}$ the mean value of the magnetization inside a cluster (which is assumed to be independent on the cluster $\mu$) and $\delta = \sigma^z_i - \bar{m}$ the fluctuations, up to the first order in $\delta$, $H_{\text{out}}$ can be written up to a constant in terms of a self-consistent longitudinal field
\begin{equation}
    H_{\text{out}} \approx H_{\text{mf}} = 2 \bar{m} \sum_{\substack{\mu,\nu \\ \mu \ne \nu}} \sum_{\substack{i \in \mu \\j \in \nu}} J_{ij} \ \sigma_i^z .
\end{equation}
We further simplify the mean-field Hamiltonian substituting $J_{i,j} \approx  J_{\mu \ell, \nu \ell}$
leading to a cluster mean-field Hamiltonian $H = \sum_\mu^{N_{cl}} H_\mu $ with
\begin{equation}
	H_\mu= -\left( \sum_{i < j \in \mu} J_{ij} \sigma_i^z     \sigma_j^z - h \sum_{i \in \mu} \sigma^x_i - 2 \bar{m} J_{\text{eff}} \sum_{i \in \mu} \sigma_i^z\right), 
    \label{eq:cluster-hamiltonian}
\end{equation}
with $J_{\text{eff}} = \lim_{N_{\text{cl}}\rightarrow\infty}\frac{1}{\mathcal{N}_{\alpha,L}} \sum_{n = 1}^{N_{\text{cl}}} \frac{J}{|n\ell|^\alpha}$. 
The local clusters are coupled only through the mean-field coefficient $J_\text{eff}$, hence we can take the limit $N_{cl} \to \infty$.


\emph{Numerical result}---
Let us now analyze the post-quench dynamics of the system using these approximations. We initially prepare the system with all spins polarized along the $z$ axis. At time $t = 0$ a finite transverse field $h$ is suddenly switched on.   
Our main observable under consideration is the dynamical order parameter $m^z $. 
We solve the dynamical equations by employing the fourth order Runge-Kutta method with an adaptive step for small cluster sizes  $\ell<10$ while for larger $\ell$ we use an MPS-TDVP technique with second-order integrator taking into account the time-dependent self-consistent longitudinal field. We set $L = 1\mathrm{e}7$ ($L=\infty$ in the TDVP). 
At each instant $t$ we integrate the Schr\"{o}dinger equation up to a time $T$, while the self-consistent field $\overline{m}$ is updated every $dt$. 

\begin{figure*}[t!]
    \centering
    \includegraphics[width=\textwidth]{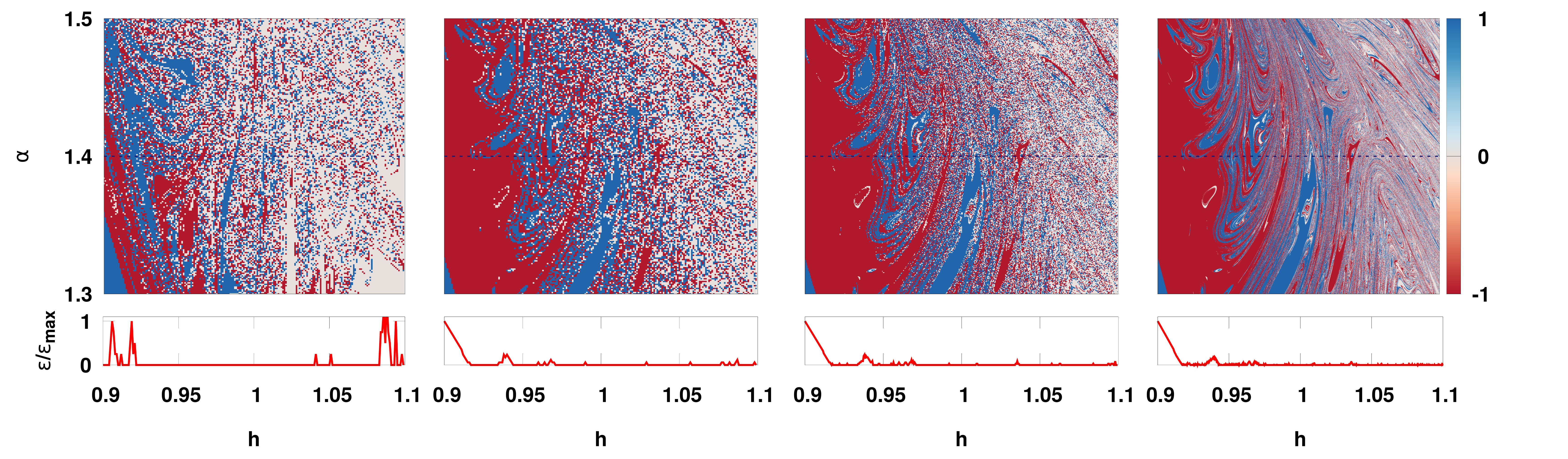}
	\caption{(Color online). A portion of the phase diagram (blue square in \fref{ddf}) for $\ell = 5$ with increasing the resolution of $\delta \alpha = \delta h = 5\mathrm{e}-3,  1\mathrm{e}-3,  5\mathrm{e}-4, 1\mathrm{e}-4$. In the critical region the phases strongly intermingle giving rise to new structures whenever the resolution is increased. Bottom panels: size of the maximum neighbourhood $\varepsilon$ containing point of the same phase evaluated for $\alpha = 1.4$ (dotted line) as a function of the post quench transverse field $h$ normalized to the valued $\varepsilon_{\text{max}} = \max_{h} \varepsilon$. Independently on the resolution $\varepsilon/\varepsilon_\text{max}$ shrinks to zero for increasing $h$. }
    \label{Chaos1} 
\end{figure*}

In \fref{ddf}, we show the dynamical phase diagram for the long-range Ising chain obtained using CMFT. It shows the sign of $m^z$ (color map) as a function of $\alpha $ and of the final value of $h$.
The phase diagram is obtained with $T = 100J$,  $dt = 1\mathrm{e}-3$ and a cluster size $\ell = 5$ which, as we will show later, despite being small can qualitatively capture the behaviour of the system. The resolution on the two axes is $\delta \alpha = 1\mathrm{e}-2$ and $\delta h = 1\mathrm{e}-3$.
The dynamical order parameter is evaluated by averaging the longitudinal magnetization in the time window $t =[80J, 100J]$.

The line $\alpha = 1$ divides the phase diagram into two different regions:
When $\alpha < 1$, as expected, the phase diagram is trivial, and we recover  the mean-field dynamical quantum phase transition at $h = 1$
~\footnote{Very close to the dynamical critical point some spurious ferromagnetic (both negative and positive) points appear. This strange behavior is just numerical noise explained by the fact that close to $h = 1$ the frequency of the oscillations of the order parameter slows down and the period becomes larger than the simulation time we chose.}.
At the mean-field level ($\ell=1$) this transition would be sharp also for $\alpha>1$.
However, accounting for short-range correlations by increasing cluster size makes the dynamical critical point spread out in a dynamical critical region that exhibits hypersensitivity of the dynamical order parameter to the model details, revealed by the alternation of points with positive and negative magnetization (as in Ref.~\cite{dqpt:4} we call this \it chaotic region \rm). The chaotic region appears already with clusters with $\ell=2$ in the phase diagram, suggesting that it is strictly connected to the presence of short-range correlations. Despite this, the form of its boundaries appears to stabilize only for larger cluster sizes  ($\ell \ge 5$). The nature of our approach limits its applicability to power law exponent $\alpha < 1.6$ (\cite{rg:1, rg:2, rg:3, rg:4}) and makes it impossible to explore the crossover at $\alpha = 2$, which is expected to be towards a short-range type behaviour where the dynamical phase transition disappears altogether.

\emph{Characterizing the chaotic region}-- The dynamics within the chaotic region is extremely rich. As shown in \fref{Chaos1}, moving from the ferromagnetic to the paramagnetic region for $\alpha>1$, the phase diagram becomes increasingly complex and the time-averaged magnetization alternates between positive and negative values for smaller and smaller changes of $h$.  As we move towards the critical point, the energy of the initial state increases and the relaxation to the stationary state becomes slower (see discussion below). The increased sensitivity is analogous to what happens when we toss a coin~\cite{coin:1}. The longer it takes the coin to release its energy the more chaotic is the system. Strzalgo and collaborators~\cite{coin:1} have in particular shown that the nature of the process can be encoded in the geometry of the phase diagram. Whenever the coin displays a trivial dynamics, the phases (head or tail) are well separated.
Conversely, in the chaotic regime, the phases intermingle leading to the formation of new structures and to the \textit{fractalization} of the phase diagram. This can be captured by defining the maximum size of the neighborhood $\varepsilon$ of points in the parameter space of the same phase. This way, chaos is defined by the condition $\varepsilon = 0$.

A fractalization of the phase diagram is clearly observed in \fref{Chaos1}. It displays  the emergence of non-trivial structures which are revealed by performing simulations with higher and higher resolution  (from the left to the right:  $\delta \alpha = \delta h = 5\mathrm{e}{-3}, 1\mathrm{e}{-3}, 5\mathrm{e}{-4}, 1\mathrm{e}{-4}$) within the chaotic region of the phase diagram ($1.3 < \alpha < 1.5$ and $0.9 < h < 1.1$).
In order to be more quantitative, we have studied the behavior of the maximum neighborhood size $\varepsilon (h)$ (bottom panels). For a fixed value $\alpha = 1.4$ (dotted line), $\varepsilon(h)$ is the size of the biggest square centered in $h$ containing points of the same colour.  
To make a comparison between the data with different resolutions possible, we consider the normalized quantity $\tilde{\varepsilon}(h) = \varepsilon(h)/\varepsilon_\text{max}$, with $\varepsilon_{\text{max}} = \max_{h} \varepsilon(h)$.
The bottom panels of \fref{Chaos1} show the behavior of $\tilde{\varepsilon}(h)$ as a function of the post-quench transverse field $h$ for the four different resolutions. As expected, it takes a maximal value in the ferromagnetic region and then shrinks to zero in the chaotic phase, independently on the resolution. 

\begin{figure}[h!]
    \centering
    \includegraphics[width=0.42\textwidth]{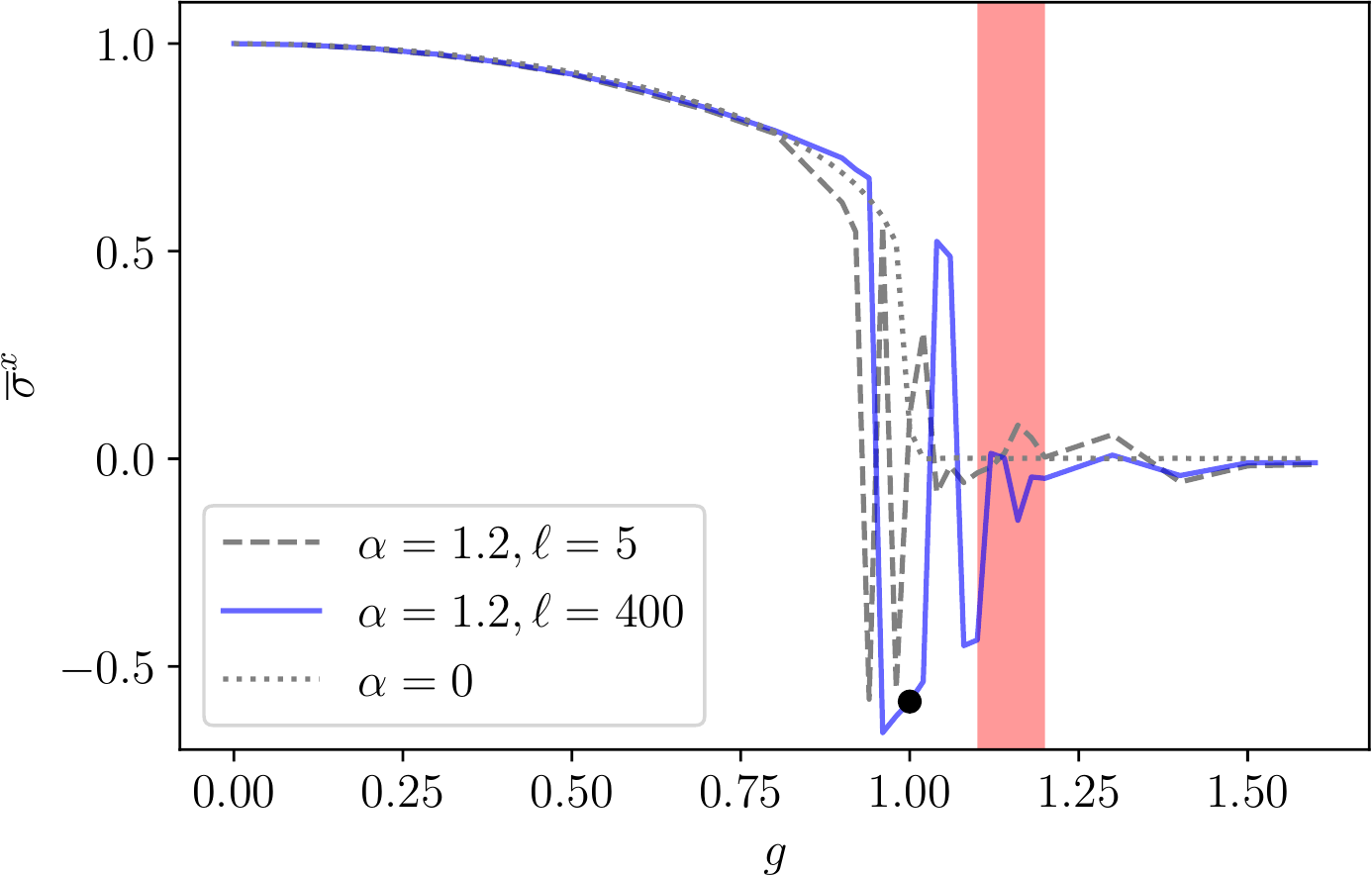}
    \caption{A cut of the dynamical phase diagram for the long-range transverse field Ising model with $\alpha=1.2$ obtained by CMFT combined with TDVP. We show the order parameter as a function of the final transverse field. At small final fields, the dynamical order parameter is close to the mean field value (gray dashed line). Close to $h=1$ we observe that the final magnetization becomes very sensitive to the quench parameter and alternates between positive and negative values.
	At large transverse fields, the dynamical order parameter vanishes. The data are obtained with the bond dimension $D=64$. The region in red denotes the range of parameters for which a stable magnetization plateau is attained at times longer than the one accessible by the MPS-TDVP.}
    \label{fig:pd slice}
\end{figure}

A feature that immediately captures the eye in \fref{Chaos1} is the presence of white, paramagnetic points within the chaotic region, whose density increases the more $h$ is increased towards the paramagnetic phase. As hinted by the fact that increasing either the cluster size (from $5$ to $7$) or the simulation time makes the density of paramagnetic points decrease systematically (see Supplementary Material), these white points are to be interpreted as points that did not settle yet in a stationary state. This is expected since the relaxation is slower deeply in the chaotic region (see Supplementary material).
 At the same time, however, this observation suggests that deep in the chaotic region the phase diagram obtained for $\ell=5$ should be taken with a grain of salt as one may easily deduce by studying the convergence of individual points to their stationary state (up or down magnetization).  

\begin{figure}[h!!]
    \includegraphics[width=0.42\textwidth]{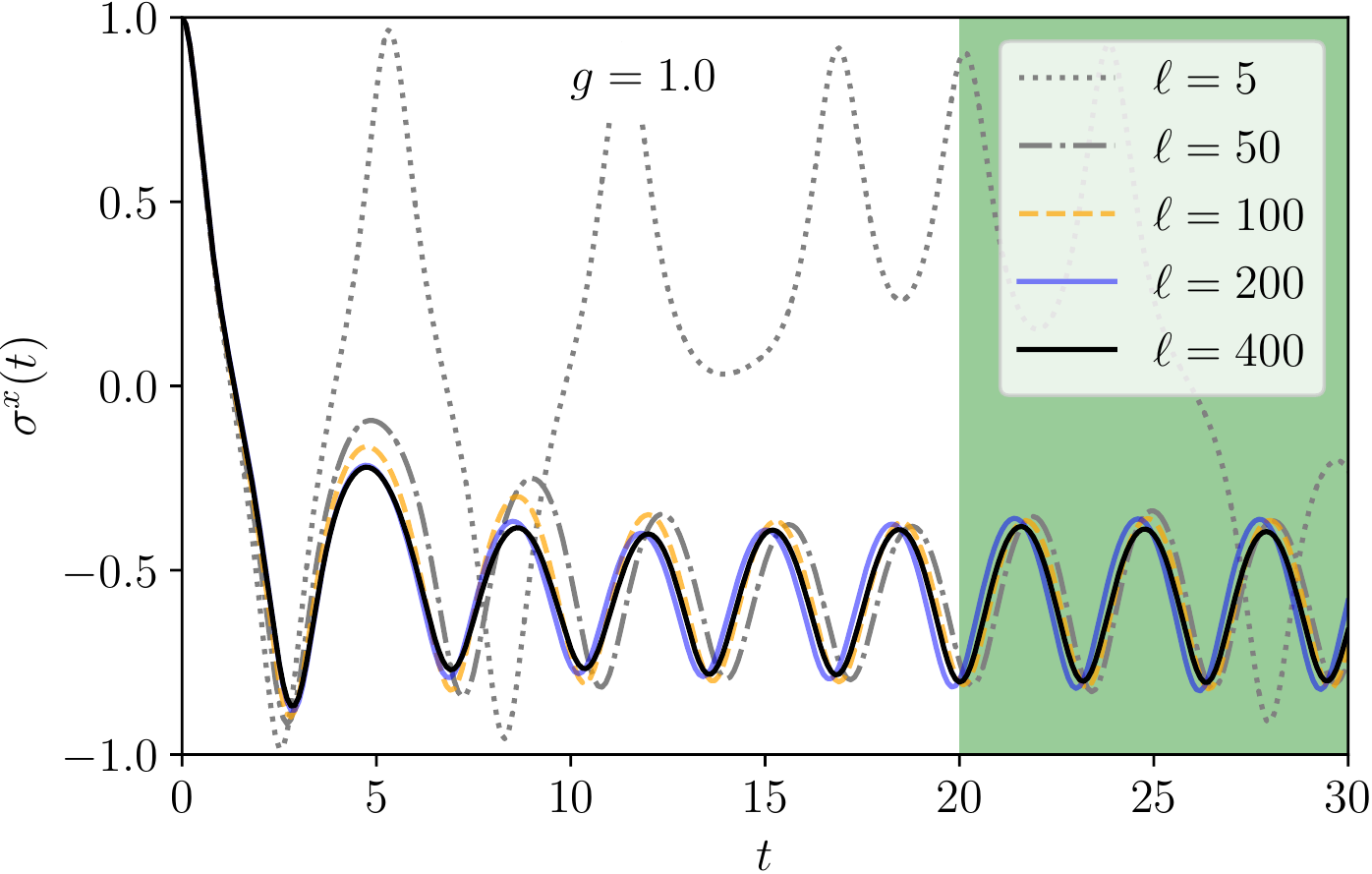}
    \caption{(Color online). Convergence of the time-dependent longitudinal magnetization $m_z(t)$ with the cluster size in the dynamical critical region at $\alpha=1.2$. Relatively large system sizes (around $\ell=200$, D = 64) are necessary to observe convergence. The order parameter $\overline{\sigma}^z$ in \fref{fig:pd slice} is obtained by averaging over the green colored region.}
    \label{fig:convergence}
\end{figure}

In the stable phases (ferromagnetic and paramagnetic), the fluctuations are weak and the final value of the order parameter is very close to the mean-field one, hence convergence is reached even with small cluster size. Instead, in the critical region, bigger cluster sizes are necessary to observe trajectory collapse. To investigate the convergence, we have used an MPS-TDVP with a second-order integrator. These results are summarized in \fref{fig:pd slice} where we display the dynamical order parameter as a function of the transverse magnetic field $h$ and in \fref{fig:convergence} where we show the convergence of a representative trajectory. The presented data is converged with  bond dimension 64 and cluster size $\ell=200$ (unless specified otherwise). We calculate the dynamical order parameter as a time average of the time-dependent longitudinal field in the time window $t=[20J,30J]$ (significantly smaller than the one used for \fref{ddf}).
Starting at small transverse fields, as expected, we observe a fast convergence of the dynamical order parameter (already at $\ell=5$).  Upon increasing the transverse magnetic field, in the vicinity of the mean field dynamical critical point, we observe a region with interchangeably positive and negative values of $m^z$. In this region, trajectories converge for relatively large cluster sizes $200\leq\ell\leq400$ (see \fref{fig:convergence}) and are sensitive to the final magnetic field. At large transverse fields, in the paramagnetic region, we again observe very fast convergence with cluster size.

Although we can reliably assess a large portion of the dynamical chaotic region, a small portion close to the transition to the paramagnetic phase remains elusive. The reasons are slower convergence with the system size and slower relaxation to a long-lived state with a well defined dynamical order parameter. As for the phase diagram in \fref{ddf}, we expect that in this region the trajectories to become even more sensitive to the control parameter and the initial condition.
As shown in \fref{fig:pd slice} the large-cluster-size simulations ($\ell=400$) agree well with the small-cluster-size simulations ($\ell=5$) also very close to the dynamical chaotic region. Therefore, these data suggest that the boundaries of the dynamical chaotic region can be estimated by simulations with small cluster sizes. Also, the tendency of $\varepsilon(h)$ to decrease in the chaotic region for increasing $h$, discussed above for $\ell=5$ is qualitatively observed in all cases, as demonstrated by studying $\varepsilon(h)$ as the cluster size is increased (see Supplementary material).

\emph{Conclusions}---
We employed a \textit{cluster mean field} theory to study the dynamics of an Ising spin chain with power-law decaying interactions. We numerically calculated the post-quench phase diagram showing that short-range correlations play an important role in determining the dynamics of this system. For some values of the power law exponent, the system exhibits an irregular dynamics that resemble the one of a tossed coin. 
Since the experimental exploration of the chaotic phase seems feasible in the near future with ion traps, we think that an investigation of the universal properties of this new phase would be intriguing.

\emph{Acknowledgement}
We thank A. Lerose and A. Gambassi for discussions.  B. \v{Z}. acknowledges support by the
Advanced grant of European Research Council (ERC),
No. 694544 – OMNES.

\bibliographystyle{ieeetr}
\bibliography{tesi}
\clearpage

\begin{appendix}
    \section{Supplementary material}
In this appendix, we further corroborate the numerical results discussed in the main text. First, we discuss the convergence of the CMFT with the cluster size. Then we discuss the stability of the chaotic dynamical region with respect to cluster size, and, finally, we study the relationship between the semiclassical energy and the chaotic behavior of the time-dependent order parameter.  

    \begin{figure}[h!]
        \centering
        \includegraphics[width = 0.51\textwidth]{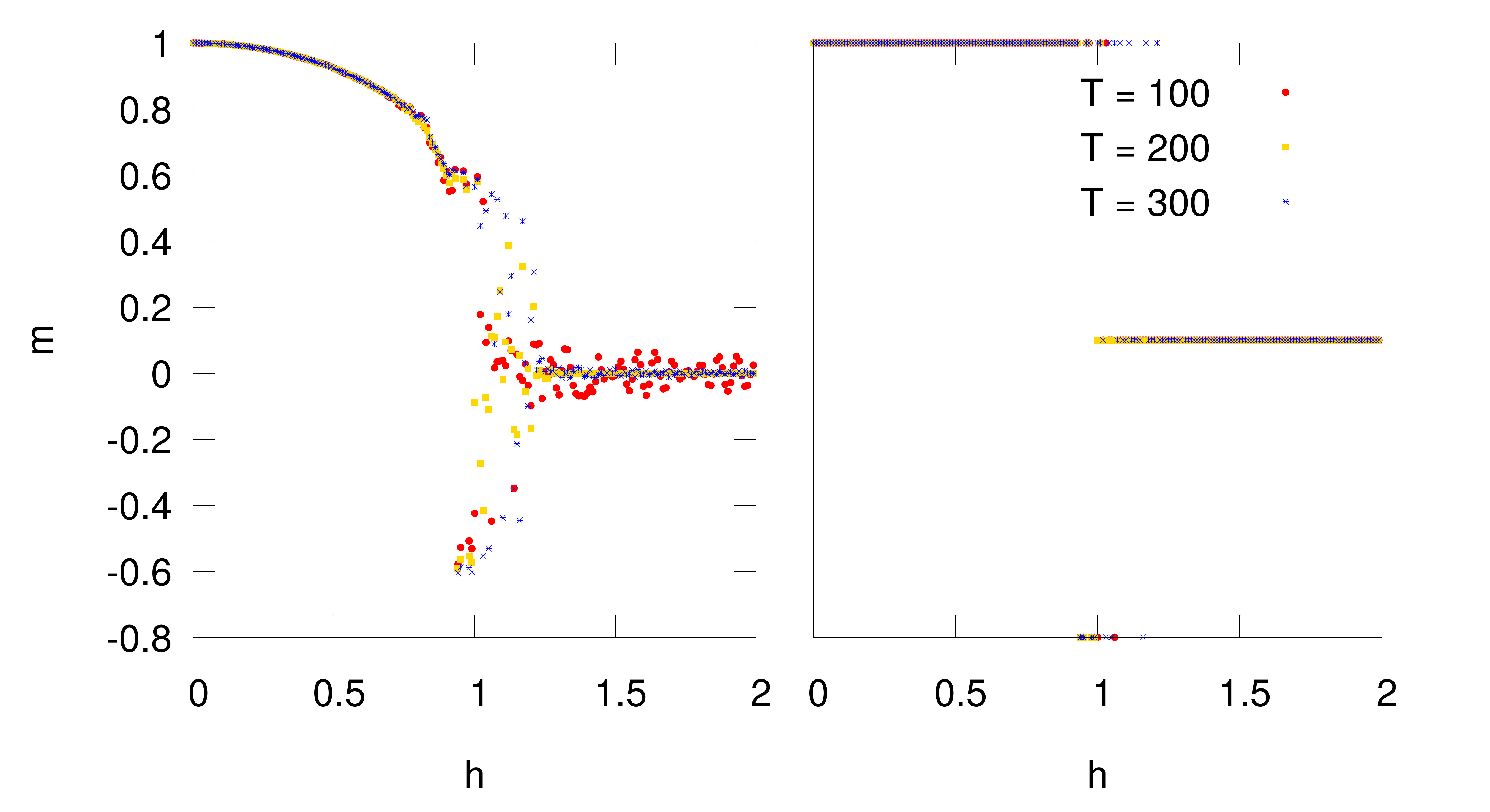}
		\caption{Phase diagram for  $\alpha = 1.2$. In the left panel we plot the order parameter $m^z$ as a function of the transverse field $h$, in the right panel the sign of the latter for the three different simulation times $T$: $T = 100$ (red points), $T = 200$ (yellow points), $T = 300$ (blue points). The averages are evaluated in the three different time windows $t \in [80, 100], [110, 200], [200, 300]$.} 
        \label{ddfc}
    \end{figure}

    \section{Convergence and finite size scaling}
    Let us investigate the data convergence both with time and cluster size. As said in the main text, the convergence of the cluster mean-field method is subtle. In the stable phase, it is reached with a relatively small cluster in a relatively small time.  Instead, in the chaotic phase, the time and the cluster sizes required are strongly dependent on the initial conditions.
	In the main text, we claim that the larger the cluster, the faster the convergence is reached suggesting that data obtained with small cluster size can be converged only by considering very long simulation times. In \fref{ddfc} we show the phase diagram for  $\alpha = 1.2$. We plot the order parameter $m^z$ (left panel) and its sign (right panel) both as a function of the post-quench transverse field $h$ for the three different simulation times $T = 100$ (red points), $T = 200$ (yellow points), $T = 300$ (blue points). The time averages are performed respectively in the time windows $t \in [80, 100], [110, 200], [200, 300]$.
    \begin{figure}[h!]
        \centering
        \includegraphics[width = 0.4\textwidth]{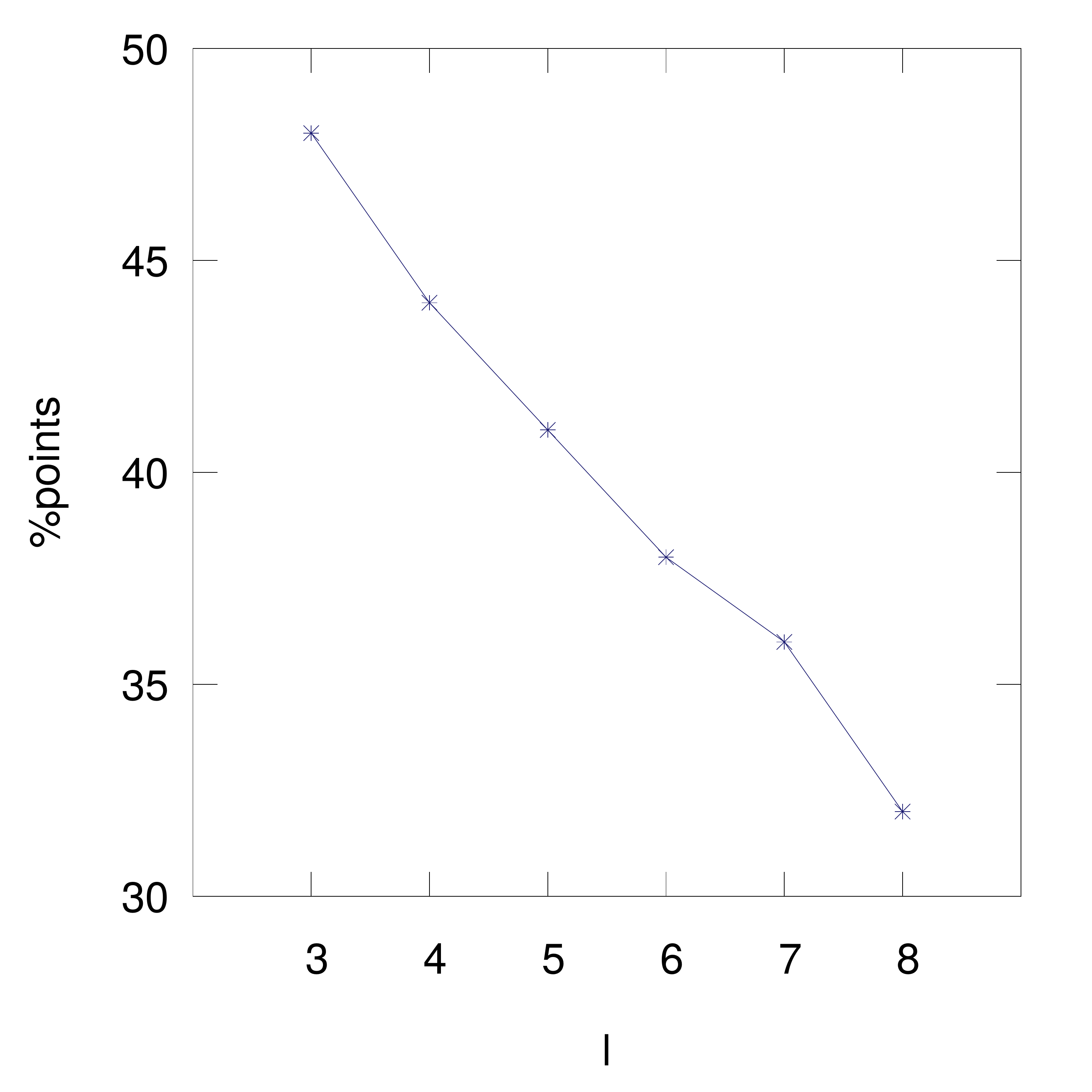}
        \caption{Percentage of these white spots in the portion of the phase diagram in \fref{ddf} whit $1 < \alpha < 1.6$ and $0.8 < h < 1.2$ as a function of the cluster length fixed the simulation time $T = 100$. This percentage is slowing down increasing cluster size suggesting a total convergence increasing $\ell$. }
        \label{bianco}
    \end{figure}
    \begin{figure*}[t!]
        \centering
        \includegraphics[width =\textwidth]{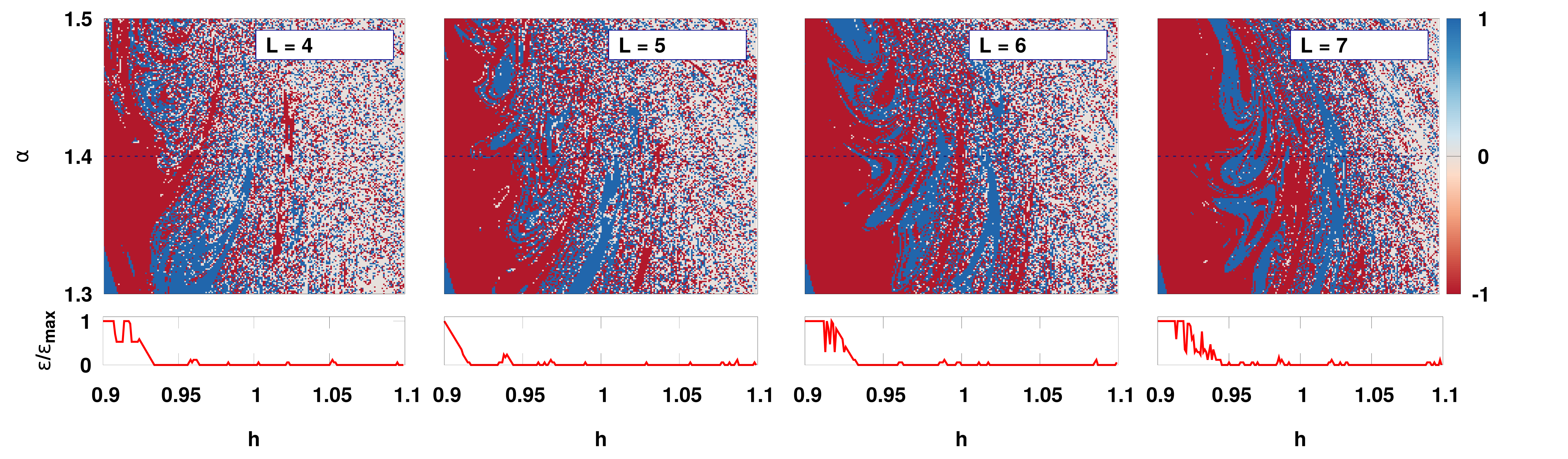}
		\caption{(Color online). This figure shows a portion (blue square in \fref{ddf} of the phase diagram (top panels) obtained with different cluster lengths (from left to right $\ell = 4, 5, 6,7$) and the relative maximum neighbourhood $\varepsilon$ of points of the same phase (bottom panels) for $\alpha = 1.4$ (dotted line) normalized to $\varepsilon_{\text{max}} = \max_{h} \varepsilon$. It turns out that, independently on the cluster size the neighbourhood shrinks to zero while moving toward the chaotic region suggesting that the fractalization process is not just a finite size effect. }
        \label{caos_L}
    \end{figure*}
    
    A good quantity that can give information about the convergence, as explained in the main text, is the percentage of spurious paramagnetic points in the phase diagram. It emerges that in the case of the simulations with $\ell = 5$ and $T = 100$ the percentage of these white spots in the portion of the phase diagram in \fref{ddf} with $1 < \alpha < 1.6$ and $0.8 < h < 1.2$ is the $41\%$ of the total area. This percentage reduces to $39\%$ increasing the simulation time up to $T = 300$. A faster decreasing can be observed by increasing the cluster size, as shown in \fref{bianco}. In this figure, the percentage of the paramagnetic points is plotted as a function of the cluster length fixed the simulation time $T = 100$. It emerges that this percentage reduces while increasing $\ell$ supporting the thesis of the total convergence in the limit of an infinite cluster. 

	Finally, we have checked that the behavior of the maximum neighbourhood $\varepsilon$ of points of the same phase does not change by increasing the cluster size. In the top panel of \fref{caos_L} we show a portion of the phase diagram (blue square in \fref{ddf}) with $dh = d\alpha = 1 \mathrm{e}-3$ for four different cluster sizes (from left to right $\ell = 4, 5, 6, 7$). In the bottom panel we plot the normalized $\varepsilon$ evaluated at $\alpha = 1.4$. We observe that $\varepsilon$ shrinks to zero independently on the cluster size, indicating that the fractalization process is not just a finite size effect. 
    \begin{figure*}[h!]
        \includegraphics[width=0.45 \textwidth]{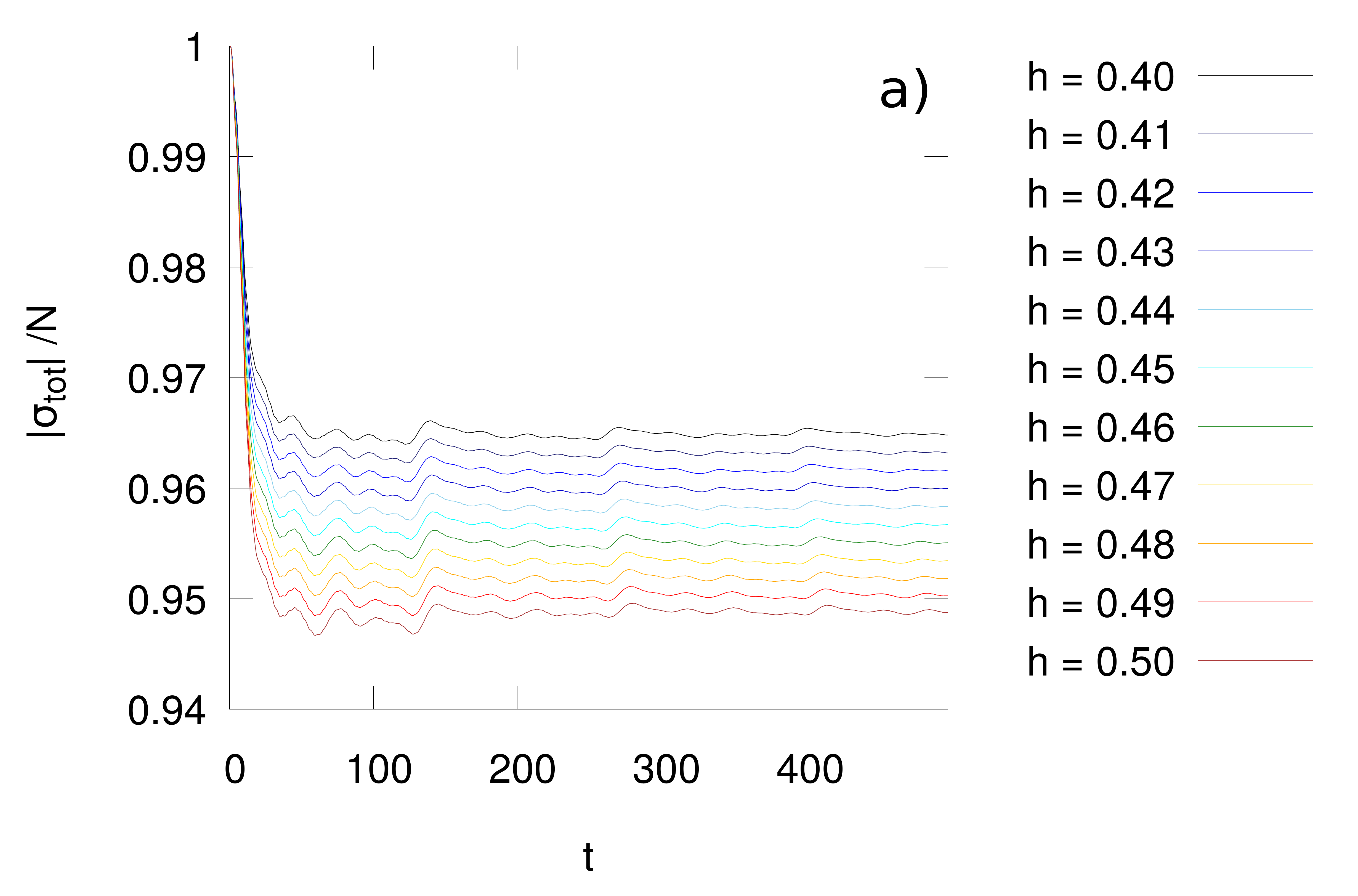}
        \includegraphics[width=0.45 \textwidth]{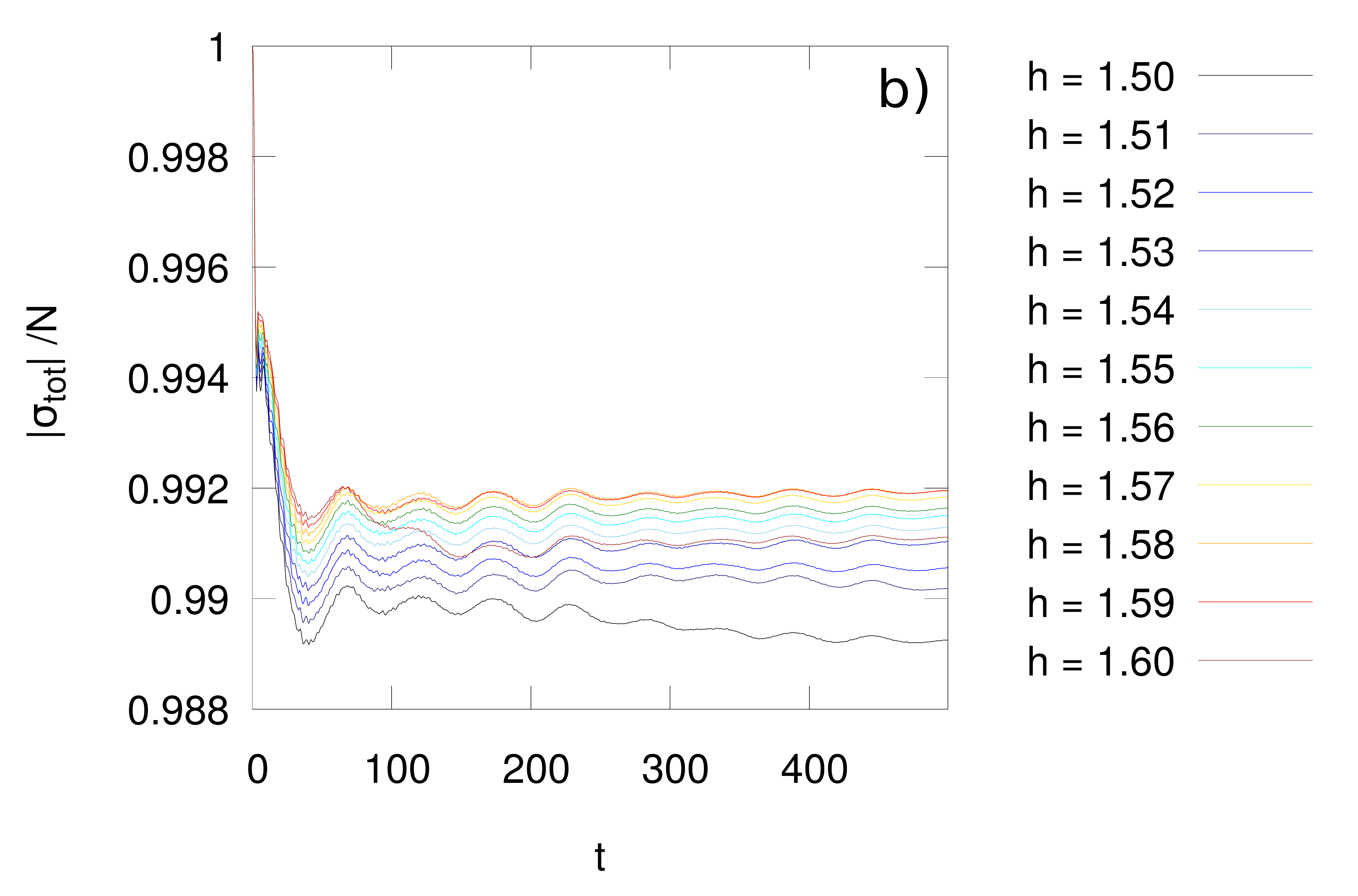}
        \includegraphics[width=0.45 \textwidth]{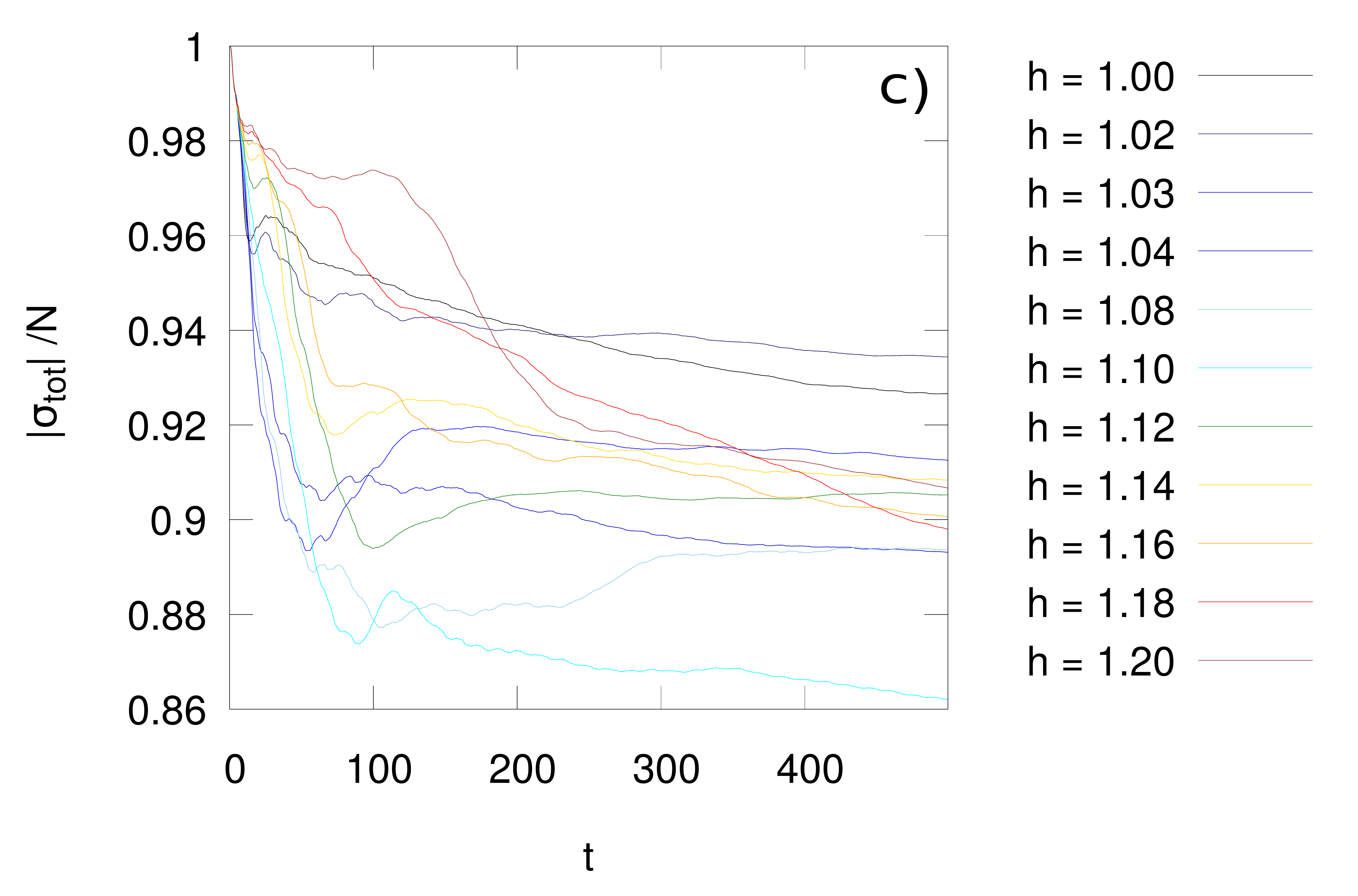}
        \caption{We show $S^2$ as a function of time in the three different phases. In the left and right panels we plot respectively the ferromagnetic and the paramagnetic phases and it is evident that the energy loss happens only in the first instants of the evolution and a typical behavior can be extract.
        In the central panel, we can see the chaotic phase. In this case, the loss of energy seems to be random and no time scale emerges.}
        \label{mod} 
    \end{figure*} 
    \section{Semiclassical energy}\label{sec:energy}
    Let us characterize the dynamic of the chaotic region from an energetic perspective in analogy with that  of a tossed coin. 
    In the pure mean field model with $\alpha < 1$ the system is described in terms of total spin ~$ \sigma_{\text{tot}} = ( \sum_i \sigma^x_i, \sum_i \sigma^y_i, \sum_i \sigma^z_i )$, with $ |\sigma_\text{tot}| = N$. Since the system is isolated, the post-quench dynamics reduces to a precession of the collective degree of freedom on the surface of the Bloch sphere. 
    For $\alpha>1$ the system can be described as a mean field collective degree of freedom with small quantum fluctuations whose role is to introduce an extensive number of microscopic degrees of freedom (the spin waves) that can scatter with the classical mode, providing new dissipation processes that concur to reduce the magnitude of the collective spin, giving rise to a nontrivial dynamics we have shown. 

We characterize these scattering processes is to study the behavior of $ \frac{|\sigma_\text{tot}|}{N}$ as a function of time in the different phases.
    The results obtained for $\alpha = 1.2$ and $\ell = 5$ are presented in \fref{mod} that shows $|\sigma_\text{tot}|/N$ as a function of time for different final transverse fields in the three phases. In the stable phases, i.e., the ferromagnetic (panel a) and the paramagnetic (panel b) one, we observe a sudden reduction of the spin modulus, sign of inelastic collisions between the classical mode and spin waves. The case of the chaotic phase (panel c) shows a more complex trend denoting that the scattering with spin waves is not solely inelastic and energy is lost after more and more collisions: the higher the number of collisions, the longer the time needed to reach the stationary state.  
    This behavior is reminiscent of the dynamics of the coin we have described in the main text. Whenever the collision with the ground is inelastic, we can observe a trivial dynamics which leads to a predictable outcome whereas we observe true randomness whenever the coin can bounce and the more the number of bounces the more the time needed to reach the final state. 
\end{appendix}

\end{document}